\documentclass[showpacs,preprint,amsmath,amssymb,aps]{revtex4}

\usepackage{graphicx}% Include figure files

\begin{document}

\def\erf{\mathrm{erf}}

\title{Spinodal instabilities of asymmetric nuclear matter within 
the Brueckner--Hartree--Fock approach}

\author{Isaac Vida\~na and Artur Polls}

\affiliation{Departament d'Estructura i Constituents de la Mat\`eria and
             Institut de Ci\`{e}ncies del Cosmos,
    Universitat de Barcelona, Avda. Diagonal 647, E-08028 Barcelona, Spain}

\begin{abstract}

We study the spinodal instabilities of asymmetric nuclear matter at finite temperature
within the microscopic Brueckner--Hartree--Fock (BHF) approximation using the realistic 
Argonne V18 nucleon-nucleon potential plus a three-body force of Urbana type. Our results 
are compared with those obtained with the Skyrme force SLy230a and the relativistic mean 
field models NL3 and TW. We find that BHF predicts a larger spinodal region. This result 
is a direct consequence of the fact that our Brueckner calculation predicts a larger critical 
temperature and saturation density of symmetric nuclear matter than the Skyrme and relativistic
mean field ones. We find that the instability is always dominated by total density fluctuations,
in agreement with previous results of other authors. We study also the restoration of the isospin
symmetry in the liquid phase, {\it i.e.,} the so-called isospin distillation or fragmentation effect,
finding that its efficiency increases with increasing proton fraction and decreases as temperature
and density increase. In general, we find that the Brueckner results are comparable to those 
obtained with the Skyrme and the relativistic mean field models, although the restoration of
isospin symmetry is not so efficient in this case.

\end{abstract}

\vspace{0.5cm}
\pacs{21.65.-f, 21.30.-x, 25.70.Pq, 24.60.Ky}

\maketitle

%%%%%%%%%%%%%%%%%%%%%%%%%%%%%%%%%%%%%%%%%%%%%%%%%%%%%%%%%%%%%%%%%%%%%%%%%%%%%%%%%%%%%%%%%%%%%%%%%%%%%%%%%%

The nature of the nucleon-nucleon interaction gives rise to a nuclear matter equation of state of Van der Waals
type and, therefore, one expects a liquid-gas like phase transition to take place \cite{ber83}.  
This transition can be explored in heavy-ion collisions; in particular the experimental results related to
the multi-fragmentation phenomenon, where highly excited composed nuclei are formed in a gas of evaporated
particles, can be interpreted as the coexistence of a liquid and a gas phase \cite{bo95,bo04,gu04,ri05,tr05}.
Since nucleons can be either neutrons or protons, one should consider nuclear matter as composed of two different 
fluids. Therefore the phase transition may lead to richer phase diagrams in the case of asymmetric nuclear matter 
\cite{barran,serot,baran98,baran01,baran05}. Phase transitions are related to the thermodynamical
instabilities that a system can present. It has been usually argued that asymmetric nuclear matter presents two 
independent types of instabilities: a mechanical (or isoscalar) instability associated to density fluctuations
which conserves the proton fraction and a chemical (or isovector) instability, related to fluctuations in the proton
fraction, occurring at constant density. Nevertheless, it has recently been shown \cite{baran01,margue03,cho04} 
that actually asymmetric nuclear matter presents only one type of instability whose direction is dominated by total 
density fluctuations which lead to a liquid-gas phase separation with restoration of the isospin symmetry in the 
liquid dense phase. This is known as the isospin distillation or fragmentation effect \cite{xu00}, where large
droplets of high density symmetric matter are formed in a background of a neutron gas with a small fraction of 
protons.

There have been systematic analysis of the stability conditions of asymmetric nuclear matter against the 
liquid-gas phase transition  by using different approaches, mean field calculations with effective 
forces of the Skyrme or Gogny type \cite{margue03, ducoin1,ducoin2,ducoin3},  relativistic mean 
field calculations using constant and density dependent coupling parameters \cite{liu02,avan04,const03,const06,const08}, 
or very recently, a Dirac--Brueckner--Hartree--Fock approach with realistic nucleon-nucleon potentials \cite{margue07}. 
In the present work, we perform a similar study starting from a realistic interaction, namely
the Argonne V18 \cite{argo} plus a three body force of Urbana type,  in the framework of the 
Brueckner--Hartree--Fock (BHF) approximation. After a brief description of the BHF approach, we will summarize the stability 
conditions and, then, discuss and compare our results with those obtained with the Skyrme force SLy230a \cite{cha97}
and the relativistic mean field models NL3 \cite{la97} and TW \cite{ty99}.

%%%%%%%%%%%%%%%%%%%%%%%%%%%%%%%%%%%%%%%%%%%%%%%%%%%%%%%%%%%%%%%%%%%%%%%%%%%%%%%%%%%%%%%%%%%%%%%%%%%%%%%%%%

Our microscopic many-body scheme starts with the construction of the $G$-matrix,
which describes the  effective  interaction between two nucleons in the presence
of a surrounding medium. It is obtained by solving the well known Bethe--Goldstone
equation
\begin{eqnarray}
\langle \vec{k}_1 \tau_1; \vec{k}_2 \tau_2 \mid G(\omega) \mid \vec{k}_3 \tau_3; \vec{k}_4 \tau_4  \rangle=
\langle \vec{k}_1 \tau_1; \vec{k}_2 \tau_2 \mid V \mid \vec{k}_3 \tau_3; \vec{k}_4 \tau_4  \rangle \nonumber \\
\displaystyle{
+\sum_{ij} 
\langle \vec{k}_1 \tau_1; \vec{k}_2 \tau_2 \mid V \mid \vec{k}_i \tau_i; \vec{k}_j \tau_j  \rangle
\frac{Q_{\tau_i\tau_j}}{\omega-E_{\tau_i}-E_{\tau_j}+i\eta}
\langle \vec{k}_i \tau_i; \vec{k}_j \tau_j \mid G(\omega) \mid \vec{k}_3 \tau_3; \vec{k}_4 \tau_4  \rangle} \ , 
\label{bg}
\end{eqnarray}
where, $\tau=n,p$, indicates the isospin projection of the two nucleons in the initial,
intermediate and final states, $\vec{k}$ are their respective linear momenta, 
$V$ denotes the bare interaction, $Q_{\tau_i\tau_j}$ is the Pauli operator which allows only intermediate
states compatible with the Pauli principle, and $\omega$, the so-called starting energy, corresponds 
to the sum of non-relativistic energies of the interacting nucleons. 
%Note the the $G$-matrices are
%obtained from a coupled channel equation. 
The single-particle energy $E_\tau$ of a nucleon with momentum
$k$ is given by
\begin{equation}
E_{\tau}(k)=\frac{\hbar^2k^2}{2m_{\tau}}+Re[U_{\tau}(k)] \ ,
\label{spe}
\end{equation}
where the single-particle potential $U_{\tau}(k)$ represents the mean field ``felt'' by a nucleon due to its
interaction with the other nucleons of the medium. In the BHF approximation, $U(k)$ is calculated through the
``on-shell energy'' $G$-matrix, and is given by
\begin{equation}
U_{\tau}(k)=\sum_{\tau'=n,p}\sum_{k'}n_{\tau'}(k') \langle \vec{k}\tau; \vec{k}'\tau' 
\mid G(\omega=E_{\tau}(k)+E_{\tau'}(k')) \mid \vec{k}\tau; \vec{k}'\tau' \rangle_A
\label{spp}
\end{equation}
where
\begin{equation}
n_{\tau}(k)=\left\{
\begin{array}{l}
1, \,\,\, $if$ \,\,  k \leq k_{F_\tau} \nonumber \\
0, \,\,\,\, $otherwise$
\end{array}
\right.
\end{equation}
is the corresponding occupation number, and the matrix elements are properly antisymmetrized. We note here
that the so-called continuous prescription has been adopted for the single-particle potential when solving
the Bethe--Goldstone equation. As shown in Refs.\ \cite{so98,ba00}, the contribution to the 
energy per particle from three-hole line diagrams is diminished in this prescription. 

All the BHF calculations carried out in this Letter use the realistic Argonne V18 \cite{argo} 
nucleon-nucleon interaction supplemented with a three-body force of Urbana type which for the use in BHF calculations 
was reduced to a two-body density dependent force by averaging over the third nucleon in the medium \cite{ba99}. 
This three-body force contains two parameters that are fixed by requiring that the BHF calculation reproduces 
the energy and saturation density of symmetric nuclear matter, see also Ref.\ \cite{zhou04} for a recent analysis
of the use of three-body forces in nuclear and neutron matter. The energy density, $E/V$, is easily 
calculated once a self-consistent solution of Eqs.\ (\ref{bg} - \ref{spp}) is achieved:
\begin{equation}
\frac{E}{V}=\frac{1}{V}\sum_{\tau=n,p}\sum_{k}n_{\tau}(k)\left(\frac{\hbar^2k^2}{2m_{\tau}}
+\frac{1}{2}Re[U_{\tau}(k)]\right) \ .
\label{ea}
\end{equation}

The many-body problem at finite temperature has been considered by several authors within different approaches,
such as the finite temperature Green's function method \cite{fe71,frick03,ar05,ar06}, the thermal  field method \cite{he95}, or the 
Bloch-De Dominicis (BD) diagrammatric expansion \cite{bl58}. The latter, represents the ``natural'' extension to finite 
temperature of the Brueckner--Bethe--Goldstone (BBG) expansion which is recovered in the zero temperature limit. Baldo 
and Ferreira \cite{ba99} showed that the dominant terms in the BD expansion were those that correspond to the zero temperature 
of the BBG diagrams, where the temperature is introduced only through the Fermi--Dirac distribution
\begin{equation}
f_{\tau}(k,T)=\frac{1}{1+exp([E_{\tau}(k,T)-\mu_{\tau}(T)]/T)} \ .
\label{fd}
\end{equation}

Therefore, at the BHF level, finite temperature effects can be introduced in a very good approximation just replacing
in the Bethe--Goldstone equation: (i) the zero temperature Pauli operator $Q_{\tau\tau'}=(1-n_{\tau})(1-n_{\tau'})$
by the corresponding finite temperature one $Q_{\tau\tau'}(T)=(1-f_{\tau})(1-f_{\tau'})$, and (ii) the single-particle 
energies $E_{\tau}(k)$ by the ones obtained when $n_{\tau}(k)$ is replaced by $f_{\tau}(k,T)$ in Eqs. (\ref{spe})
and (\ref{spp}) . In this case, the self-consistent process implies that together with the 
Bethe--Goldstone equation and the single-particle potential, the chemical potential $\mu_{\tau}(T)$ of the nucleon, must 
be extracted at each step of the iterative process, since both the Bethe--Goldstone equation and the single-particle
potential depend implicitly on it.

Once a self-consistent solution is obtained the free-energy density is determined by the usual thermodynamical relation,
\begin{equation}
{\cal F} = \frac{E}{V}-T\frac{S}{V} \ ,
\label{fa}
\end{equation}
where $E/V$ is evaluated from Eq.\ (\ref{ea}) replacing $n_{\tau}(k)$ by $f_{\tau}(k,T)$, and the entropy density is
calculated in the quasi-particle approximation
\begin{equation}
\frac{S}{V}=-\frac{1}{V}\sum_{\tau}\sum_{k}[f_{\tau}(k,T)ln(f_{\tau}(k,T))+(1-f_{\tau}(k,T))ln(1-f_{\tau}(k,T))] \ .
\label{sa}
\end{equation}

We have now all the necessary ingredients to analyze the stability of asymmetric nuclear matter against phase 
separation. The stability of the system is guaranteed if the free-energy of a single phase is lower than the 
free-energy in any two-phases configurations. This condition is fulfilled if the free-energy density is a convex 
function of the neutron and proton densities 
$\rho_n$ and $\rho_p$, that is, if the curvature matrix,
\cite{serot,baran98,margue03}
\begin{equation}
{\cal C}_{ij}  = \left ( \frac {\partial ^2 {\cal F}} {\partial \rho_i \partial \rho_j} \right )_T, \ \ \ i,j=n,p
\end{equation}
is positive-definite. This demands that both its trace and its determinant should be positive,
{\it i.e.,}
\begin{equation}
Tr({\cal C}_{ij})  = \lambda_+ + \lambda_- \geq 0, \ \ \ Det({\cal C}_{ij})=\lambda_+\lambda_- \geq 0 \ ,
\end{equation}
where
\begin{equation}
\lambda_{\pm} = \frac {1}{2}  \left ( Tr ({\cal C}_{ij}) \pm  {\sqrt {Tr( {\cal C}_{ij})^2 - 4 Det ( {\cal C}_{ij})}}\right )
\end{equation}
are the two real eigenvalues of the curvature matrix associated to the eigenvectors $(\delta\rho^{\pm}_n,\delta\rho^{\pm}_p)$ 
\begin{equation}
\frac {\delta \rho_i^{\pm}}{\delta \rho_j^{\pm} } = \frac {\lambda_{\pm} - {\cal C}_{jj}}{{\cal C}_{ji}}, \ \ \  i,j = p,n \ .
\end{equation}
Stability requires, therefore, that both eigenvalues should be positive. In asymmetric nuclear matter $\lambda_+$ is always 
positive, and only $\lambda_-$ can eventually become negative, signaling the beginning of the instability and the phase 
separation \cite{margue03}. It turns out that the magnitude of $\lambda_+$ exceeds always that of $\lambda_-$ 
($\lambda_+ > |\lambda_-|$) and therefore the trace appears to be always positive. This is due, as it has been already pointed 
out by Margueron and Chomaz in Ref.\ \cite{margue03}, to the fact that: (i) the dominant dependence of the energy on 
the isospin asymmetry, $\beta=(\rho_n-\rho_p)/(\rho_n+\rho_p)$, is essentially quadratic, and (ii) the symmetry energy 
is a positive increasing function of the total density, $\rho=\rho_n+\rho_p$. Consequently, the spinodal region at a given 
temperature will be just determined by the values of neutron and proton densities which make the determinant of the curvature 
matrix negative. 

Before presenting our results, we will give some technical details concerning the evaluation of the second derivatives of
the free-energy density needed to construct the curvature matrix.  As it has already been pointed out, the dominant dependence
of the energy on the isospin asymmetry is essentially quadratic, therefore, at a given temperature, one can separate in 
good approximation the density and isospin asymmetry dependence of the free-energy density assuming that
\begin{equation}
{\cal F}(\rho,\beta) \approx {\cal F}_0(\rho)+{\cal F}_{sym}(\rho)\beta^2 \ ,
\label{eq:parabol}
\end{equation}
where ${\cal F}_0(\rho)={\cal F}(\rho,0)$ is the free-energy density of symmetric nuclear matter, and 
${\cal F}_{sym}={\cal F}(\rho,1)-{\cal F}(\rho,0)$. Now, expressing all the derivatives with respect to neutron
and proton densities in terms of derivatives with respect to $\rho$ and $\beta$, and assuming
Eq.\ (\ref{eq:parabol}), one can perform all the derivatives with respect to $\beta$ analytically and write the 
matrix elements of the curvature matrix as a function only of ${\cal F}_0(\rho)$, ${\cal F}_{sym}$, and their
first and second order derivatives with respect to $\rho$,
\begin{equation}
{\cal C}_{nn}={\cal F''}_0(\rho)+{\cal F''}_{sym}(\rho)\beta^2+\frac{4\beta (1-\beta)}{\rho}{\cal F'}_{sym}(\rho) 
+\frac{6\beta^2-8\beta+2}{\rho^2}{\cal F}_{sym}(\rho) \ , 
\label{eq:matelnn}
\end{equation}
\begin{equation}
{\cal C}_{np}={\cal C}_{pn}={\cal F''}_0(\rho)+{\cal F''}_{sym}(\rho)\beta^2-\frac{4\beta^2}{\rho}{\cal F'}_{sym}(\rho) 
+\frac{6\beta^2-2}{\rho^2}{\cal F}_{sym}(\rho)  \ ,
\label{eq:matelnp}
\end{equation}
\begin{equation}
{\cal C}_{pp}={\cal F''}_0(\rho)+{\cal F''}_{sym}(\rho)\beta^2-\frac{4\beta (1+\beta)}{\rho}{\cal F'}_{sym}(\rho) 
+\frac{6\beta^2+8\beta+2}{\rho^2}{\cal F}_{sym}(\rho) \ .
\label{eq:matelpp}
\end{equation}
The derivatives ${\cal F''}_0(\rho), {\cal F'}_{sym}(\rho)$ and ${\cal F''}_{sym}(\rho)$ can also be done analytically
if one assumes a particular functional form of the free-energy density in symmetric and neutron matter. In our case,
we have chosen to fit the quantity $\rho \ d(F/A)/d\rho$ for symmetric and  neutron matter and to express all the derivatives of
${\cal F}_0(\rho)$ and ${\cal F}_{sym}(\rho)$ with respect to $\rho$ in terms of this quantity and its derivatives. The reason
for such a choice is double, first $\rho \ d(F/A)/d\rho$ goes in both cases, symmetric and neutron matter, to the temperature
in the limit of very low densities, so one has a fix point in the fit, and second, this choice turns out to be numerically very stable.

%%%%%%%%%%%%%%%%%%%%%%%%%%%%%%%%%%%%%%%%%%%%%%%%%%%%%%%%%%%%%%%%%%%%%%%%%%%%%%%%%%%%%%%%%%%%%%%%%%%%%%%%%%%%%%%

\begin{figure}[thb]
   \includegraphics[height=8.5cm,angle=0]{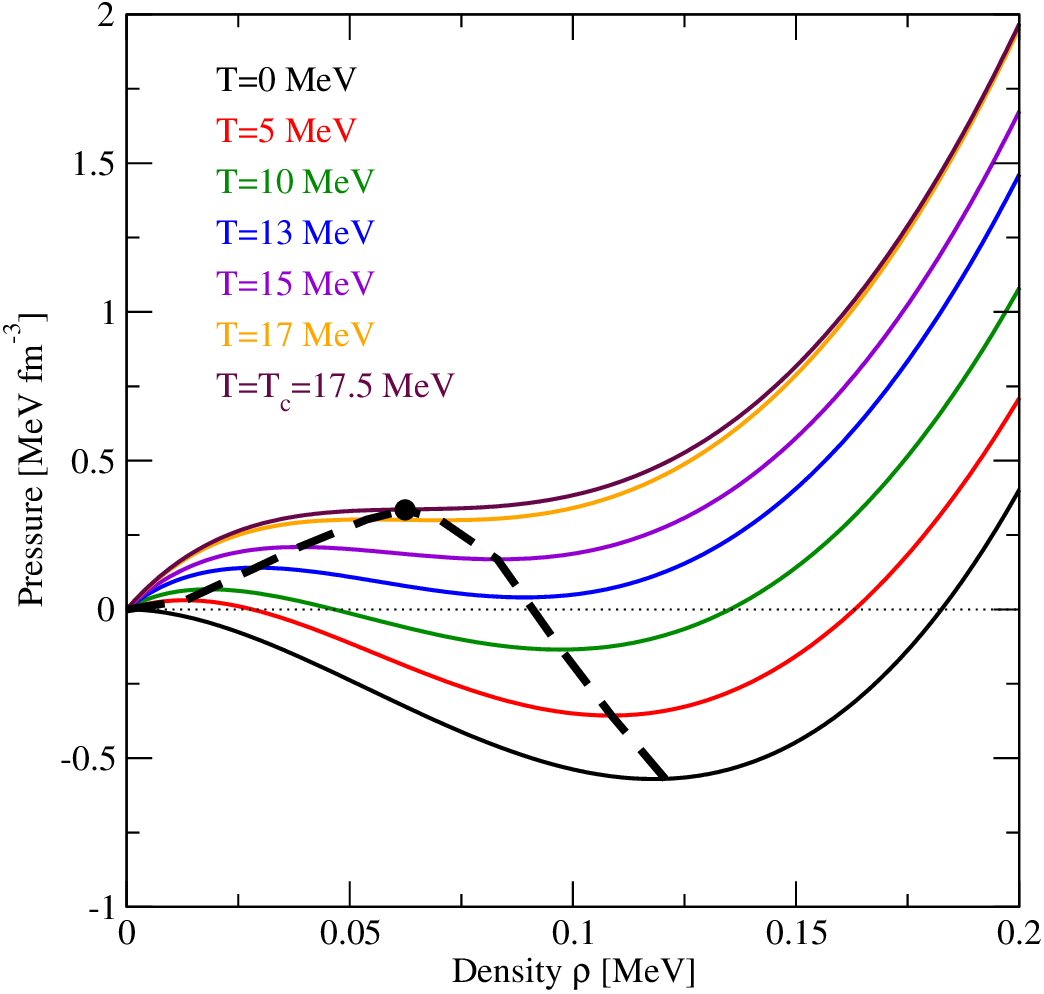}
   \vspace{0.25cm}
   \caption{(Color online) Symmetric nuclear matter isotherms for several temperatures. The border of the spinodal zone is delineated by
the black dashed curve; uniform matter below this curve is mechanically unstable. The critical point, indicated by the solid dot,
corresponds to a temperature of $T_c=17.5$ MeV.}
   \label{fig:fig1}
\end{figure}

In the next we present the results of our microscopic Brueckner approach, and compare them with those obtained with the 
Skyrme force SLy230a and the relativistic mean field models NL3 and TW. We start by showing in Fig.\ \ref{fig:fig1}
symmetric nuclear matter isotherms obtained in our BHF approach with Argonne V18 plus a Urbana type three body force
for several temperatures ranging from $T=0$ up to the critical temperature $T_c=17.5$ MeV (indicated by the solid
dot). The border of the spinodal zone is delineated by the black dashed curve. As it has been already said, the
spinodal region is determined by the values of $\rho_n$ and $\rho_p$ for which Det$({\cal C}_{ij})<0$. In symmetric
nuclear matter this region is exactly the one where the pressure decreases with density, and therefore where the system
is mechanically unstable. As a general feature, we can observe that the spinodal instability starts at zero density, temperature
and pressure, and disappears well before the saturation density $\rho_0=0.182$ fm$^{-3}$ is reached.

\begin{figure}[thb]
   \includegraphics[height=7.5cm,angle=0]{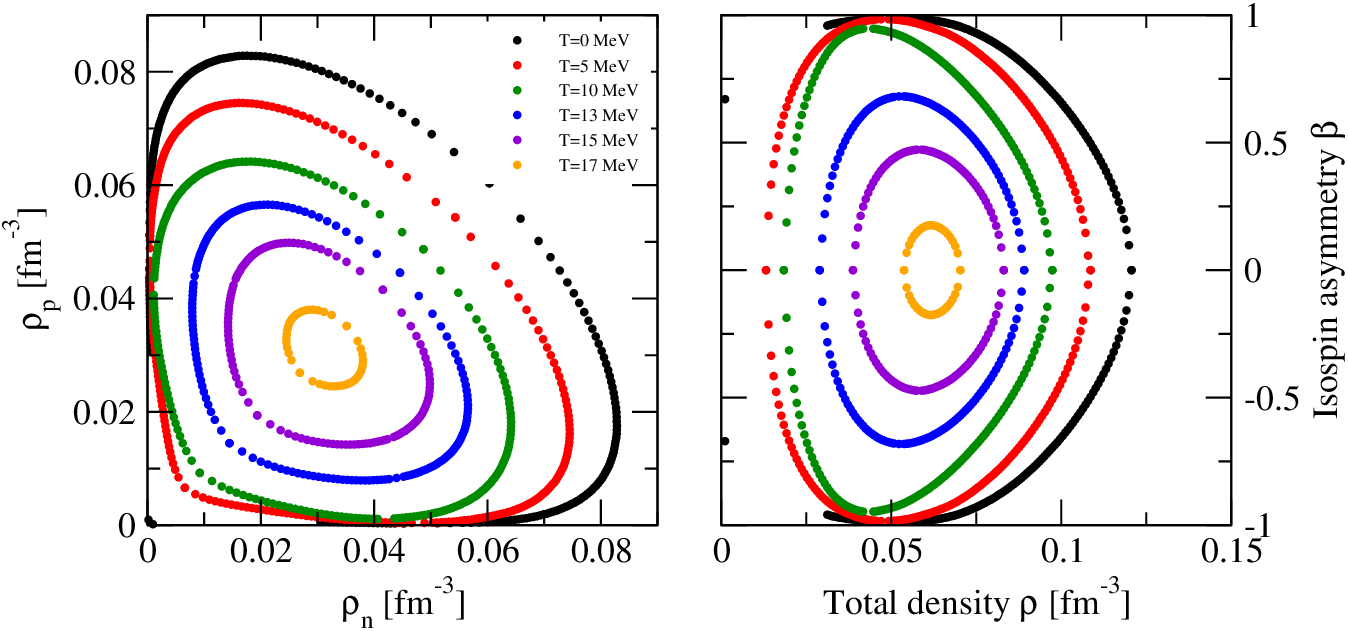}
   \vspace{0.25cm}
   \caption{(Color online) Projection of the spinodal contour in the neutron-proton density
plane (left panel) and in the density-$\beta$ plane (right panel) for temperatures ranging
from  $T=0$ up to $T=17$ MeV.} 
   \label{fig:fig2}
\end{figure}

Let us consider now the most general case of asymmetric nuclear matter. Fig.\ \ref{fig:fig2} depicts
the projection of the spinodal contours in the neutron-proton density plane (left panel) and in the density-$\beta$ plane (right panel)
for several temperatures ranging from $T=0$ up to $T=17$ MeV. The qualitative behaviour is very similar to the one observed in previous 
calculations using different relativistic or non relativistic mean field models 
\cite{margue03,cho04,xu00,ducoin1,ducoin2,ducoin3,liu02,avan04,const03,const06,const08}. As the temperature increases, the instability region, 
defined by the inner part of the spinodal boundaries, shrinks up to the point corresponding to the critical 
temperature of symmetric nuclear matter $T_c$. Note that the isospin symmetry of the nucleon-nucleon interaction 
ensures that the contours are symmetric respect to the interchange of $\rho_n$ and $\rho_p$ (left panel)
or the interchange of $\beta$ and $-\beta$ (right panel). Notice also (see right panel) that for a given temperature, 
the total density at which the spinodal instability appears decreases when the isospin asymmetry increases, similar 
to what happens with some Skyrme (SLy230a, SGII) and Gogny (D1P) forces \cite{margue03}. We have also considered
the presence of a degenerated ultrarelativistic free Fermi gas of electrons in the system with a density equal to that of the 
protons in order to keep charge neutrality. In this case, the curvature matrix ${\cal C}$ becomes 
\begin{equation}
{\cal C}=\left(
\begin{array}{l}
\partial^2 {\cal F}/\partial \rho_n^2 \ \ \ \ \ \ \ \ \ \ \ \   \partial^2 {\cal F}/\partial \rho_n \partial \rho_p  \\ 
\partial^2 {\cal F}/\partial \rho_p \partial \rho_n \ \ \partial^2 {\cal F}/\partial \rho_p^2 + \partial \mu_e/\partial \rho_e 
\end{array}
\right) \ ,
\end{equation}
where $\mu_e$ is the electron chemical potential. We have found that in the density region of interest for our study, the electron 
contribution to the curvature matrix, $\partial \mu_e/\partial \rho_e$, is so large that it leads to a complete suppression of the 
instability region already at zero temperature. Nevertheless, this result is model dependent. For example, in Ref.\
\cite{ducoin3}, using Skyrme forces SLy230a, SGII and SIII, it was also found a complete suppression of the instability region, while for 
instance in Ref.\ \cite{const03}, where the relativistic mean field model NL3 was used, it is reported a strong reduction of 
the instability region in the case of nuclear matter with electrons, but not its complete suppression.

\begin{figure}[thb]
   \includegraphics[height=8.5cm,angle=0]{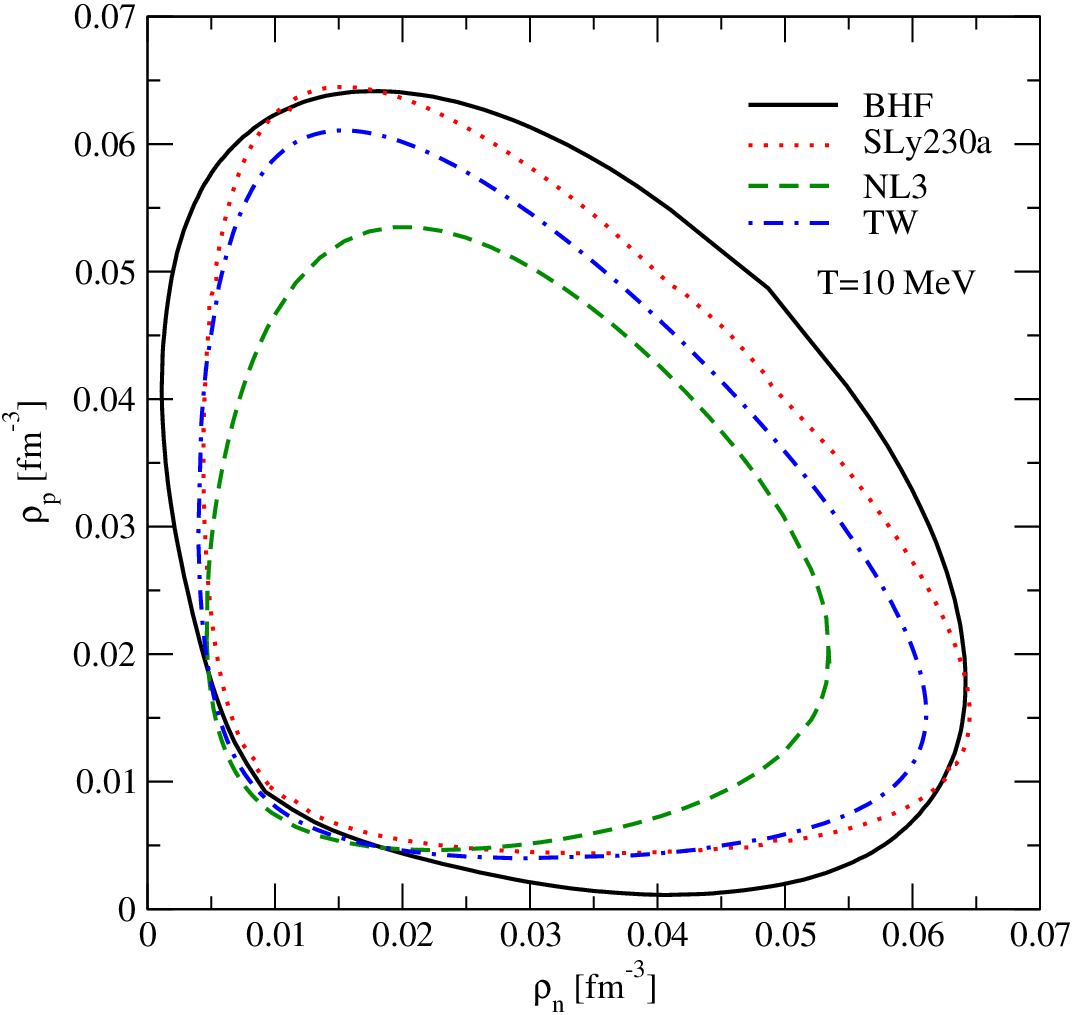}
   \vspace{0.25cm}
   \caption{(Color online) Comparison of the spinodal contour at $T=10$ MeV predicted by 
our BHF approach (black solid line) with those obtained by us with the Skyrme force SLy230a (red dotted line)
and those obtained by Avancini {\it et al.} \cite{const06} using the relativistic meand 
field models NL3 (green dashed line) and TW (blue dot-dashed line).} 
   \label{fig:fig3}
\end{figure}

Let us compare now our Brueckner prediction for the spinodal region with those obtained with the Skyrme force SLy230a 
and the relativistic mean field models NL3 and TW. This comparison is presented in Fig.\ \ref{fig:fig3} for a fixed 
temperature of $10$ MeV. The results for the Skyrme force SLy230a have been calculated also by us, whereas those
of the relativistic mean field models NL3 and TW have been taken from Ref.\ \cite{const06}.
Clearly the larger spinodal region, which almost envolves all the others, corresponds to the Brueckner one. 
This can be easily understood noting that in the symmetric nuclear matter case the larger the 
critical temperature and saturation density at zero temperature are, the larger the spinodal region is. 
Although in asymmetric nuclear matter the onset of the instability region will be sensitive to the density dependence of the
symmetry energy and its derivatives, one can still expect that the model that predicts the larger critical temperature 
and saturation density for symmetric matter will predict the larger spinodal instability region in asymmetric matter. 
In fact, our Brueckner calculation predicts the larger critical temperature, $T_c=17.5$ MeV and saturation density, $\rho_0= 0.182$ fm$^{-3}$.
The closest curve to the Brueckner prediction corresponds to the one obtained for the non-relativistic Skyrme force SLy230a,
which predicts a critical temperature of $T_c=14.6$ MeV and a saturation density $\rho_0= 0.160$ fm$^{-3}$. Finally, the critical
temperatures for NL3 and TW are $14.5$ and $15.1$ MeV respectively, while the saturation densities are quite similar, $0.148$
fm$^{-3}$ for NL3 and 0.153 fm$^{-3}$ for TW, predicting smaller spinodal regions. 

Usually, it has been argued that asymmetric nuclear matter presents two independent types of instabilities: a mechanical 
(or isoscalar) instability conserving the proton fraction and a chemical (or isovector) instability occurring at constant density. 
Nevertheless, recently Margueron and Chomaz \cite{margue03} (see also Refs.\ \cite{baran01} and \cite{cho04}) have shown that 
in fact instabilities in asymmetric nuclear matter appear as a mixture of density and proton fraction fluctuations, and, therefore, 
they cannot be separated into mechanical or chemical instabilities: chemical (mechanical) instabilities do not only involve changes in the 
proton fraction (density) but also changes in the density (proton fraction). There is only one type of instability whose direction 
is given by the eigenvector $(\delta\rho^-_n,\delta\rho^-_p)$ associated with the negative eigenvalue $\lambda_-$ \cite{baran01,margue03,cho04}. 
If $\delta\rho^-_p/\delta\rho^-_n=\rho_p/\rho_n$ then the instability preserves the ratio between protons and neutrons at which the
system was prepared, and its nature is purely mechanical, whereas if $\delta\rho^-_p=-\delta\rho^-_n$ then the total density remains constant and 
therefore the instability is purely chemical.  In general, however, the nature of the instability will never be neither purely mechanical
nor purely chemical, but it will appear as a mixture of both which will be predominantly of isoscalar 
type if $\delta\rho^-_p/\delta\rho^-_n >0$ or of isovector type if $\delta\rho^-_p/\delta\rho^-_n<0$ \cite{cho04,baran01,margue03}. 

\begin{figure}[thb]
   \includegraphics[height=8.5cm,angle=0]{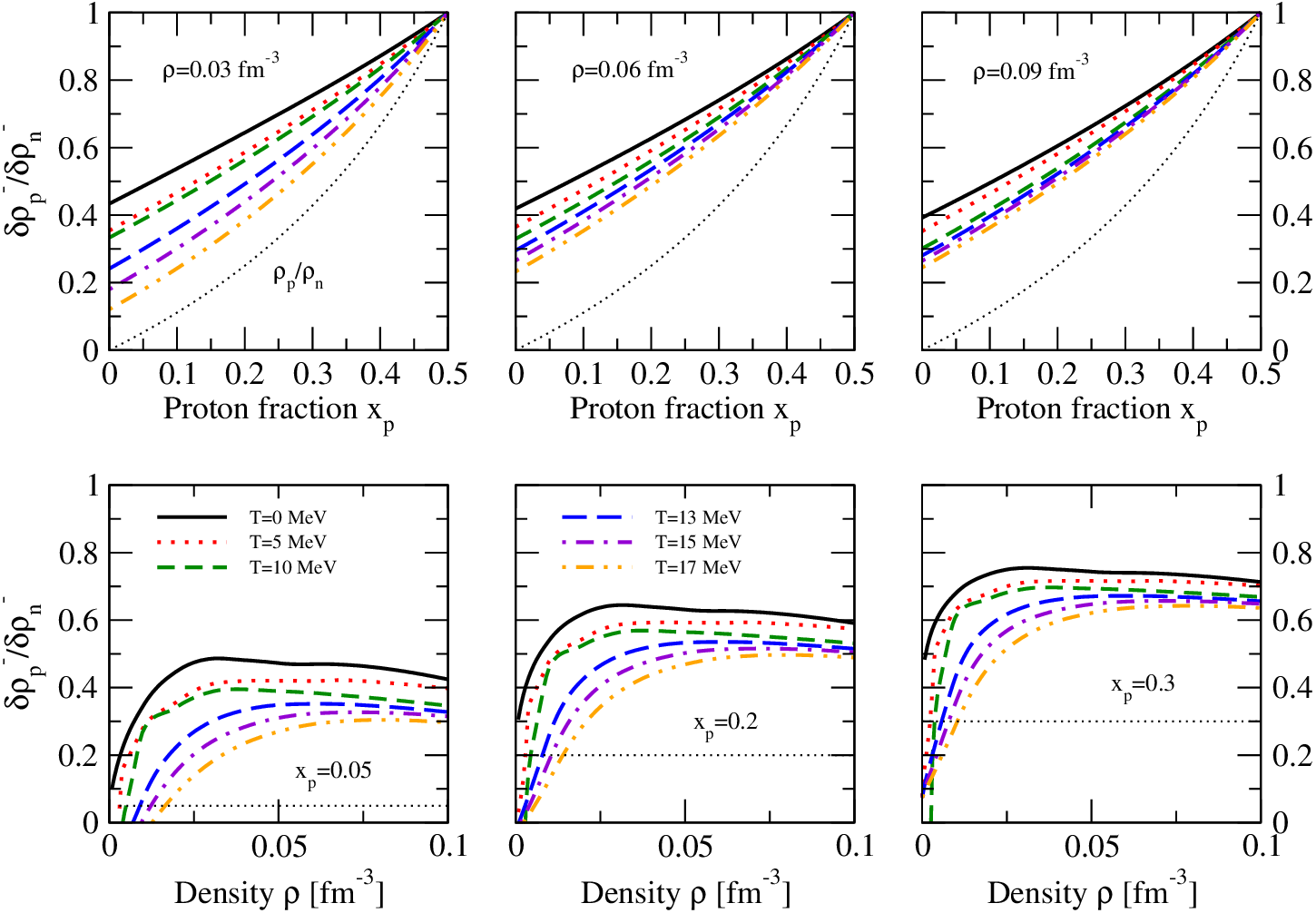}
   \vspace{0.25cm}
   \caption{(Color online) Upper panels: Ratio $\delta \rho^-_p/\delta \rho^-_n $ as a function of 
the proton fraction for several temperatures and fixed densities, $\rho=0.03$ (left panel), $0.06$
(middle panel) and $0.09$ (right panel) fm$^{-3}$. Black dotted line shows the ratio between protons
and neutrons $\rho_p/\rho_n$. Lower panels: Same ratio as a function of the total density for the 
same set of temperatures and fixed proton fractions, $x_p= 0.05$ (left panel), $x_p=0.2$ (middle panel)
and $x_p=0.3$ (right panel).}
   \label{fig:fig4}
\end{figure}

Figure \ref{fig:fig4} shows the ratio $\delta \rho^-_p/\delta \rho^-_n$ as a function of the proton fraction (upper panels) for
several temperatures ranging from $T=0$ up to $T=17$ MeV and three fixed densities: $\rho=0.03$ (left panel), $0.06$ (middle panel) 
and $0.09$ (right panel) fm$^{-3}$. The same ratio is showed in the lower panels as a function of the total density for the same 
set of temperatures and proton fractions: $x_p= 0.05$ (left panel), $x_p=0.2$ (middle panel) and $x_p=0.3$ (right panel).
Note that in all cases $\delta \rho^-_p/\delta \rho^-_n$ is positive, indicating that the instability is always dominated by 
the total density fluctuations ({\it i.e.,} it is of isoscalar type) even for large asymmetries, in agreement with the 
results of Refs. \cite{baran01}, \cite{margue03} and \cite{const06}. This is expected because the symmetry energy in our approach is always
a positive increasing function of density, and therefore we do not expect the instabilities to be dominated by proton fraction
fluctuations.  In particular, we notice that for symmetric matter ($x_p=0.5$) $\delta \rho_p^-/ \delta \rho_n^- = 1$, meaning 
that in this case the instability occurs in the pure isoscalar direction and that matter behaves as a one component system. 
Moreover, it can be observed that $\delta\rho^-_p/\delta\rho^-_n$
is larger than the ratio between protons and neutrons $\rho_p/\rho_n$ (see upper panels), which tells us that the instability 
drives the dense phase (liquid) of the system towards a more isospin symmetric region of the $\rho_n$-$\rho_p$ plane. The 
conservation of the total particle number enforces the light phase (gas) to become more neutron rich leading 
to the so-called isospin distillation or fragmentation effect \cite{xu00}. The ratio $\delta \rho^-_p/\delta \rho^-_n$ can be then 
considered a measurement of the efficiency of isospin symmetry restoration in the liquid phase: the larger its value the 
greater the efficiency. In general, we observe that whereas the restoration of isospin symmetry becomes more and more efficient
when the proton fraction increases ($\delta \rho_p^-/ \delta \rho_n^-$ increases), it becomes less efficient when the 
temperature and the density increase ($\delta \rho_p^-/ \delta \rho_n^-$ decreases). Only for very low densities 
($\rho \leq 0.02$ fm$^{-3}$) this efficiency grows with increasing density.

\begin{figure}[thb]
   \includegraphics[height=8.5cm,angle=0]{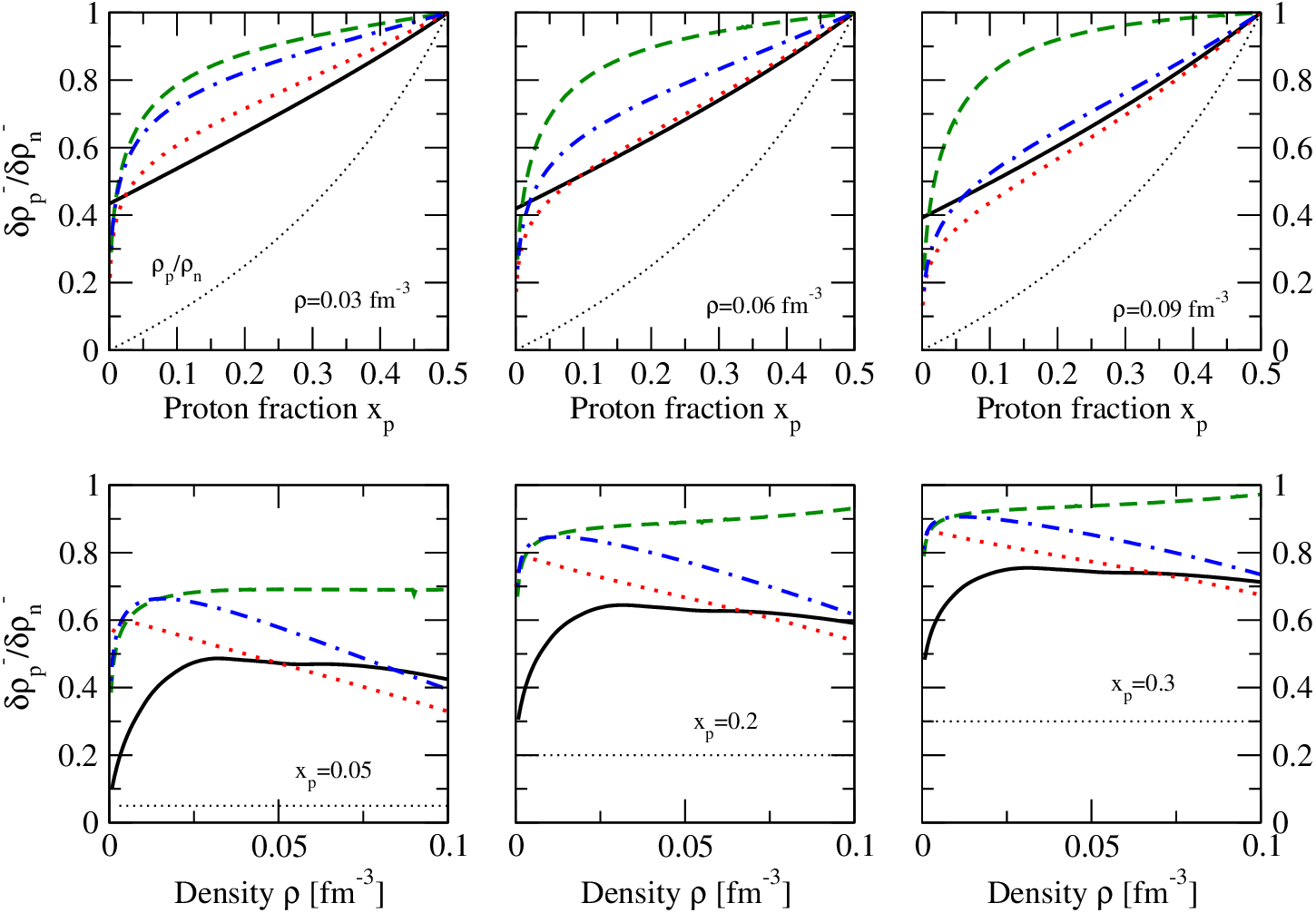}
   \vspace{0.25cm}
   \caption{(Color online) Comparison of the  ratio $\delta \rho^-_p/\delta \rho^-_n $ at $T=0$
predicted by our BHF approach (black solid lines) with those obtained by us with the Skyrme force SLy230a (red dotted lines) 
and those obtained by Avancini {\it et al.} \cite{const06} using the
relativistic mean field models NL3 (green dashed lines) and TW (blue dot-dashed lines). As in the previous figure, in the
upper panels this ratio is plotted as a function of the proton fraction for three fixed densities, 
$\rho=0.03, 0.06$ and $0.09$ fm$^{-3}$, whereas its density dependence is shown in the lower panels
for fixed proton fractions, $x_p= 0.05$ (left panel), $x_p=0.2$ (middle panel) and $x_p=0.3$ (right panel).}
   \label{fig:fig5}
\end{figure}

Finally, we compare in Fig.\ \ref{fig:fig5} our Brueckner prediction for the isospin symmetry restoration with those of the
Skyrme SLy230a force and the relativistic mean field models NL3 and TW. In general, one can say that in the Brueckner approach 
the restoration of isospin symmetry is less efficient ($\delta \rho_p^-/ \delta \rho_n^-$ is smaller) than in the case of Skyrme 
or relativistic mean field, except for very low proton fractions. We also note (see lower panels) that whereas the Brueckner
approach, the Skyrme SLy230a and the TW model share the same qualitative behaviour ($\delta \rho_p^-/ \delta \rho_n^-$ 
decreases with density), the NL3 model predicts the opposite one ($\delta \rho_p^-/ \delta \rho_n^-$ increases with density).
As it has been discussed by Avancini {\it et al.} in Ref.\ \cite{const06}, this is essentially due to the presence of the 
rearrangement contribution in the evaluation of the neutron and proton chemical potentials. This contribution is present 
in the BHF approach, in the Skyrme force and in the TW model, but it is completely absent in the NL3. In fact, the authors of 
Ref.\ \cite{const06} showed that by removing this contribution from the TW model, the results predicted by this model
(TW with no rearrangement) show a behaviour similar to NL3. 

\begin{figure}[thb]
   \includegraphics[height=8.5cm,angle=0]{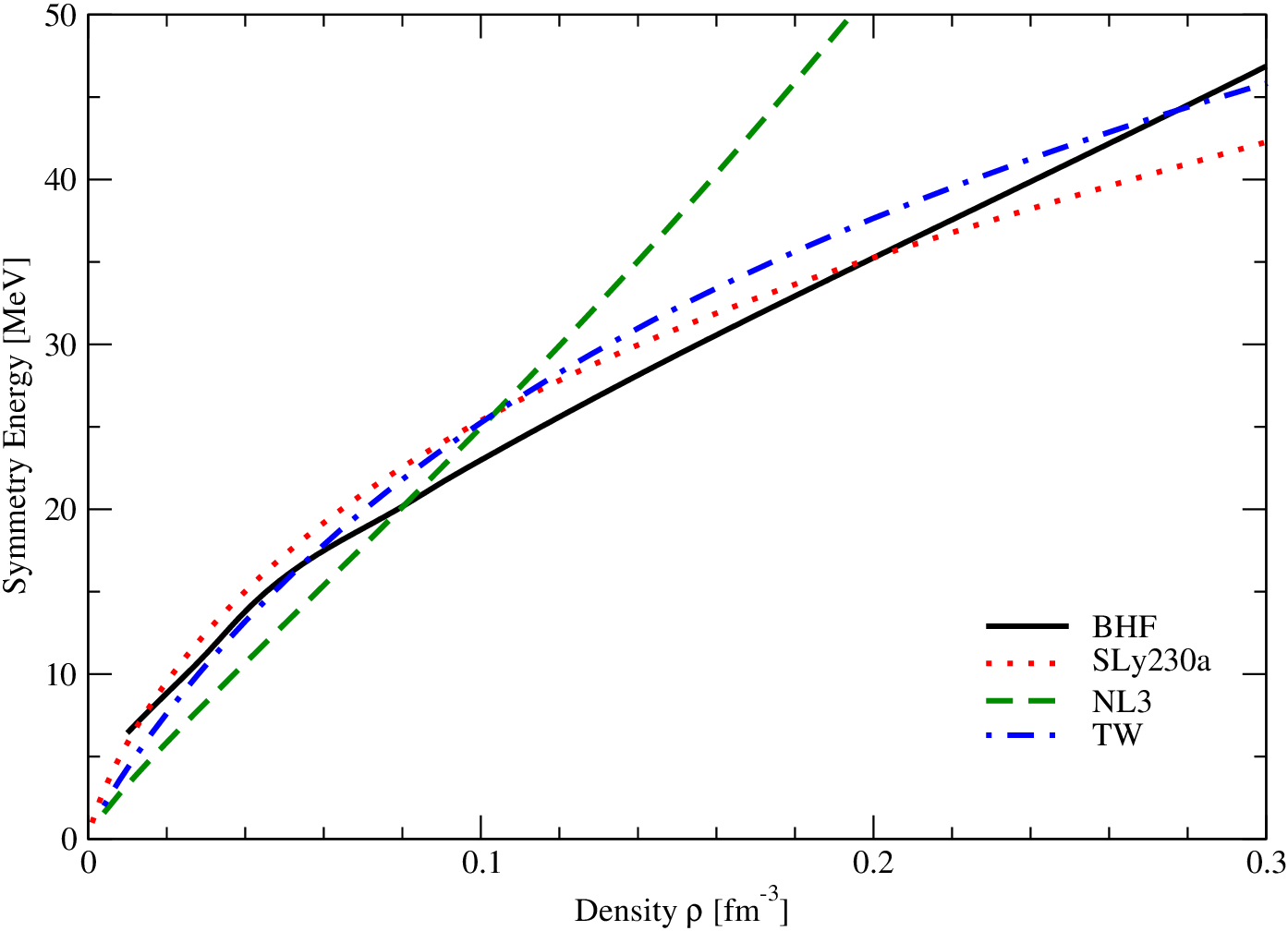}
   \vspace{0.25cm}
   \caption{(Color online) Comparison of the symmetry energy as a function of the density obtained within our
BHF approach (black solid line), with those obtained with the Skyrme force SLy230 (red dotted line) and the
relativistic mean field models NL3 (green dashed line) and TW (blue dot-dashed line).}
   \label{fig:fig6}
\end{figure}

To enlight the analysis of the distillation effect, we report in Fig.\ \ref{fig:fig6} the density dependence of the
symmetry energy predicted by the different models. Nevertheless, it is clear from Eqs.\ (\ref{eq:matelnn},\ref{eq:matelnp},\ref{eq:matelpp})
that the matrix elements of the curvature matrix, and therefore the ratio $\delta \rho_p^-/ \delta \rho_n^-$, depend not only 
on the value of the symmetry energy but also on its first and second derivatives, and therefore to draw a conclusion 
on the distillation effect for the different effective interactions is not straightforward and would require a more sophisticated 
and careful analysis, than just considering the value of the symmetry energy. This analysis is beyond the scope of the present work
and it will be addressed in the future. The figure also serves to illustrate that whereas the BHF approach, the Skyrme SLy230a force 
and the TW model predicts a qualitatively similar behaviour of the symmetry energy as a function of density, the NL3 model clearly 
shows a different one, being almost linear. This different behaviour of the symmetry energy is in agreement with the fact, already discussed,
that NL3 model predicts also a behaviour of the ratio $\delta \rho_p^-/ \delta \rho_n^-$ qualitatively different from the one predicted
by the other three models.

Summarizing, we have studied the spinodal instabilities of asymmetric nuclear matter at 
finite temperature within the microscopic Brueckner--Hartree--Fock approach using 
the realistic Argonne V18 potential plus a three-body force of Urbana type which guarantees the reproduction 
of the saturation density in symmetric nuclear matter. The results are qualitatively similar to 
previous analysis performed with non-relativistic and relativistic effective interactions. In particular,
we have compared our results with those obtained with the Skyrme force SLy230a and the relativistic mean
field models NL3 and TW, finding that the Brueckner approach predicts a larger spinodal region. This is mainly 
related to the fact that our Brueckner calculation with Argonne V18 plus a Urbana type three-body force
predicts values for the critical temperature and the saturation density of symmetric nuclear matter larger 
than the ones provided by those effective interactions.  We have found that the instability is always 
dominated by total density fluctuations, in agreement with previous results of other authors 
\cite{baran01,margue03,const06}. Finally, we have also studied the restoration of the isospin symmetry in the liquid phase, 
finding that whereas it becomes more and more efficient when proton fraction increases, it becomes less and less
efficient with the increase of the temperature and the density. In general, we found that the BHF results 
are comparable to those obtained with the Skyrme and the relativistic mean field models, 
although the restoration of isospin symmetry is not so efficient in this case.

\vspace{0.6cm}
  
We are very grateful to Constan\c{c}a Provid\^encia and J\'{e}r\^{o}me Margueron for useful and 
stimulating discussions and comments, and to Bruno Juli\'a-D\'{i}az for a careful reading of the 
manuscript. In particular, we would also like to thank Constan\c{c}a Provid\^{e}ncia for providing
us with the results of the relativistic mean field models NL3 and TW. This work is supported by 
Grant No. FIS2005-03142 (Ministerio de Educaci\'on y Ciencia) and Grant No. 2005SGR-00343
(Generalitat de Catalunya).

%%%%%%%%%%%% FIGURES %%%%%%%%%%%%%%%%%%%%%%%%%%%%

\newpage

\newpage

\newpage

\newpage

\newpage

\newpage

\end{document}